\documentclass[a4paper,12pt]{article}
\usepackage{amsthm,amsmath,amssymb,latexsym}

\newtheorem{thm}{Theorem}[section]
\newtheorem{prop}[thm]{Proposition}

\newtheorem{rem}[thm]{Remark}

\numberwithin{equation}{section}

\newcommand{\ds}{\displaystyle}

\newcommand{\vep}{\varepsilon}

\title{Classical solutions of the degenerate Garnier system
and their coalescence structures}
\author{Takao Suzuki\\ Department of Mathematics, Kobe University,\\
Rokko, Kobe 657-8501, Japan\\
{\small suzukit@math.kobe-u.ac.jp}}
\date{}

\begin{document}

\maketitle

%% Abstract
\begin{abstract}
We study the degenerate Garnier system which generalizes the fifth Painlev\'{e}
equation $P_{\rm{V}}$.
We present two classes of particular solutions, classical transcendental and
algebraic ones.
Their coalescence structure is also investigated.
\end{abstract}

%% Section 1
\section{Introduction}
The Painlev\'{e} equations $P_J$ $(J=\rm{I},\ldots,\rm{VI})$ are derived from
the theory of monodromy preserving deformations of linear differential equations
of the form
\begin{equation*}
	(L_J)\quad\frac{d^2y}{dx^2} + p_1(x,t)\,\frac{dy}{dx} + p_2(x,t)\,y = 0,
\end{equation*}
with singularities corresponding to a partition of four as follows (see e.g.
\cite{IKSY}):
\begin{equation*}
	\begin{array}{|c|c|}\hline
		L_{\rm{VI}}&(1,1,1,1)\\
		L_{\rm{V}}&(1,1,2)\\
		L_{\rm{IV}}&(1,3)\\
		L_{\rm{III}}&(2,2)\\
		L_{\rm{II}}&(4)\\ \hline
	\end{array}
\end{equation*}
In this table, a partition $(r_1,\ldots,r_k)$ indicates that $L_J$ has $k$
singularities of Poincar\'{e} ranks $r_1-1,\ldots,r_k-1$, respectively.
Thus we regard each of $P_J$ $(J=\rm{II},\ldots,\rm{VI})$ as an equation
corresponding to a partition of four.
We note that the length $k$ of the partition equals the number of constant
parameters contained in $P_J$.
The first Painlev\'{e} equation $P_{\rm{I}}$ has no constant parameter and does
not correspond to any partition.

The Garnier system (in $N$ variables) generalizes the sixth Painlev\'{e}
equation $P_{\rm{VI}}$ and governes the monodromy preserving deformation of
linear differential equation with $N+3$ regular singularities \cite{IKSY}.
We also regard the Garnier system as corresponds to the partition $(1,\ldots,1)$
of $N+3$.

Each of the Painlev\'{e} equations $P_J$ $(J=\rm{I},\ldots,\rm{V})$ can be
reduced from the sixth one through a certain limitting procedure, in parallel
with the confluence of singularities of the linear differential equation $L_J$
\cite{OKM1}.
Similarly, the degenerations of the Garnier system are considered
\cite{KAW1,KAW2,KIM,LIU,OKM3}.
Each of them is associated with a partition.
We denote by $G(r_1,\ldots,r_k;N)$ the degenerate Garnier system in $N$
variables corresponding to a partition $(r_1,\ldots,r_k)$ of $N+3$ .

It is well known that each of $P_J$ $(J=\rm{II},\ldots,\rm{VI})$ admits two
classes of classical solutions, hypergeometric and algebraic (or rational) ones.
The coalescence structure of these solutions is investigated in detail
\cite{MSD1,MSD2}, as well as the degeneration scheme of the Painlev\'{e}
equations.
Also, the Garnier system $G(1,\ldots,1;N)$ has such classes of classical
solutions \cite{KO,TSU1,TSU2,TSU3}.
The aim of this paper is to study particular solutions of the degenerate Garnier
system $G(1,\ldots,1,2;N)$ and their coalescence structure by means of
$\tau$-functions.

We have in \cite{SUZ} a family of $\tau$-functions for $G(1,\ldots,1;N)$ 
arranged on a lattice.
This family is determined by a certain completely integrable Pfaffian system.
In Section \ref{Sec:degGar}, we investigate the degeneration of the Pfaffian
system together with the degenerate limitting procedure from $G(1,\ldots,1;N)$
to $G(1,\ldots,1,2;N)$; hence we obtain a family of $\tau$-functions on a
lattice for $G(1,\ldots,1,2;N)$.
We have in particular (see Theorems \ref{Thm:CTSol}, \ref{Thm:RatSol} and
\ref{Thm:AlgSol})
\begin{thm}
The system $G(1,\ldots,1,2;N)$ admits three types of solutions{\rm:}\\[4pt]
{\rm(i)} classical transcendental ones expressed by the hypergeometric series
$\Phi_D${\rm;}\\
{\rm(ii)} rational ones in terms of the Schur polynomials{\rm;}\\
{\rm(iii)} algebraic ones in terms of the universal characters.
\end{thm}

%% Section 2
\section{Degenerate Garnier system}\label{Sec:degGar}
In this section, we formulate the degenerate Garnier system $G(1,\ldots,1,2;N)$,
then introduce a family of $\tau$-functions for the system.

%% Section 2.1
\subsection{Hamiltonian system and Schlesinger system}
Let $\{,\}$ be the Poisson bracket defined by
\begin{equation}
	\{f,g\} = \sum_{j=1}^{N}\left(\frac{\partial f}{\partial p_j}
	\frac{\partial g}{\partial q_j}-\frac{\partial g}{\partial p_j}
	\frac{\partial f}{\partial q_j}\right).
\end{equation}
Consider the following completely integrable Hamiltonian system:
\begin{equation}\label{Sys:degGar}
	dq_j = \sum_{i=1}^{N}\,\{K_i,q_j\}\,ds_i,\quad
	dp_j = \sum_{i=1}^{N}\,\{K_i,p_j\}\,ds_i\quad (j=1,\ldots,N),
\end{equation}
with polynomial Hamiltonians $K_i$ $(i=1,\ldots,N)$:
\begin{equation}
	\begin{split}
		s_1^2K_1
		&= q_1\left(\rho+\sum_{j=1}^{N}q_jp_j\right)
		\left(\rho+\theta_{N+3}+1+\sum_{j=1}^{N}q_jp_j\right)\\
		&\qquad
		+ \sum_{j=2}^{N}\,s_1p_1q_j
		- \sum_{j=2}^{N}\,s_jq_1(q_jp_j-\theta_j)p_j
		- \sum_{j=2}^{N}\,(s_j-1)q_jp_j\\
		&\qquad
		- s_1q_1p_1(q_1p_1-\theta_{N+2})+(q_1-s_1)p_1,\\[4pt]
		s_i(s_i-1)K_i
		&= q_i\left(\rho+\sum_{j=1}^{N}q_jp_j\right)
		\left(\rho+\theta_{N+3}+1+\sum_{j=1}^{N}q_jp_j\right)\\
		&\qquad
		- \sum_{j=2,j\neq i}^{N}R_{ij}\,q_ip_i(q_jp_j-\theta_j)
		- \sum_{j=2,j\neq i}^{N}R_{ji}\,q_i(q_jp_j-\theta_j)p_j\\
		&\qquad
		- \sum_{j=2,j\neq i}^{N}S_{ij}\,p_i(q_ip_i-\theta_i)q_j
		- \sum_{j=2,j\neq i}^{N}R_{ij}\,(q_ip_i-\theta_i)q_jp_j\\
		&\qquad
		+ \left\{s_ip_i-(s_i+1)q_ip_i\right\}(q_ip_i-\theta_i)
		+ (\theta_{N+2}s_i+\theta_{N+1}-1)q_ip_i\\
		&\qquad
		+ \frac{s_i(s_i-1)}{s_1}\left\{q_ip_i+p_i(q_ip_i-\theta_i)q_1\right\}
		- (s_i-1)q_ip_1\\
		&\qquad
		- s_i(2q_ip_i-\theta_i)q_1p_1\qquad (i=2,\ldots,N),
	\end{split}
\end{equation}
where
\begin{equation}
	\sum_{j=2}^{N+3}\,\theta_j+2\,\rho=0,
\end{equation}
and
\begin{equation}
	R_{ij} = \frac{s_i(s_j-1)}{s_j-s_i},\quad
	S_{ij} = \frac{s_i(s_i-1)}{s_i-s_j}.
\end{equation}
We call \eqref{Sys:degGar} the {\it degenerate Garnier system} and
denote it by $G(1,\ldots,1,2;N)$.
This system is regarded as a generalization of the fifth Painlev\'{e} equation
$P_V$ \cite{OKM2}.
For $N=1$, this is exactly the Hamiltonian system of $P_V$.
We note that $G(1,\ldots,1,2;N)$ is equivalent to the system given by H. Kimura
\cite{KIM} via a certain canonical transformation.

Let $A_j$ $(j=1,\ldots,N+2)$ be matrices of the dependent variables defined by
\begin{equation}
	A_j = \begin{pmatrix}a_j&b_j\\c_j&d_j\end{pmatrix}.
\end{equation}
Consider the following system of differential equations:
\begin{equation}\label{Sys:degSch}
	\begin{split}
		&dA_1 = \sum_{i=2}^{N+1}\,[A_i,A_1]\,d\log t_i
		+ \left(A_1+[A_{N+2},A_1]\right)d\log t_1,\\
		&dA_j = \sum_{i=2,i\neq j}^{N+2}[A_i,A_j]\,d\log\,(t_j-t_i)
		+ \frac{[A_1,A_j]}{t_j}\,d\log\frac{t_j}{t_1}\quad
		(j=2,\ldots,N+1),\\
		&dA_{N+2} = \sum_{i=2}^{N+1}\left(\frac{[A_i,A_1]}{t_i}\,
		d\log\frac{t_i}{t_1} + [A_i,A_{N+2}]\,d\log t_i\right),
	\end{split}
\end{equation}
where $t_{N+1}=1$ and $t_{N+2}=0$.
Here we assume
\begin{enumerate}
\item\quad $\mathrm{tr}A_1=t_1$,\quad
$\mathrm{tr}A_j=\theta_j\notin\mathbb{Z}$\quad $(j=2,\ldots,N+2)$;
\item\quad $\mathrm{det}A_j=0$\quad $(j=1,\ldots,N+1)$,\quad
$\mathrm{tr}A_1A_{N+2}=t_1\theta_{N+2}$;
\item\quad The matrices $A_j$ satisfy
\begin{equation}
	A_{\infty}:=-\sum_{j=2}^{N+2}\,A_j
	= \begin{pmatrix}\rho&0\\0&\rho+\theta_{N+3}\end{pmatrix},\quad
	\theta_{N+3}\notin\mathbb{Z}.
\end{equation}
\end{enumerate}
We call \eqref{Sys:degSch} the {\it degenerate Schlesinger system} denoted by
$S(1,\ldots,1,2;N)$.

The system $S(1,\ldots,1,2;N)$ is in fact equivalent to $G(1,\ldots,1,2;N)$ via
\begin{equation}
	\begin{split}
		&s_1 = -\frac{1}{t_1},\quad s_i = \frac{t_i-1}{t_i},\\
		&q_1 = -\frac{b_1}{t_1b_{\infty}},\quad
		q_i = (t_i-1)\,\frac{b_i}{b_{\infty}}\quad,\\
		&q_1p_1 = a_1 + a_{N+2} - b_1\,\frac{a_{N+1}}{b_{N+1}}
		- b_{N+2}\,\frac{a_1}{b_1},\\
		&q_ip_i = a_i - t_ib_i\,\frac{a_{N+1}}{b_{N+1}}
		+ (t_i-1)\,b_{i}\,\frac{a_1}{b_1}\quad (i=2,\ldots,N),
	\end{split}
\end{equation}
where $b_{\infty}=b_1+\sum_{j=2}^{N+2}t_jb_j$.

Recall that both of $G(1,\ldots,1,2;N)$ and $S(1,\ldots,1,2;N)$ govern the
holonomic deformation of the system of linear differential equations:
\begin{equation}\label{Eq:LDE_degSch}
	\frac{d\vec{y}}{dx} = A(x,t)\,\vec{y},\quad
	A(x,t) = \frac{A_1(t)}{x^2} + \sum_{j=2}^{N+2}\,\frac{A_j(t)}{x-t_j},
\end{equation}
concerning the parameter $t=(t_1,\ldots,t_N)$; see \cite{JMU}.

%% Section 2.2
\subsection{A family of $\tau$-functions}
\begin{prop}[\rm\cite{JM}\bf]\label{Prop:1-form_degGar}
For each solution of $S(1,\ldots,1,2;N)$, the 1-form
\begin{equation}
	\omega_0 = \sum_{i=1}^{N}\,H_i\,dt_i,
\end{equation}
is closed.
Here we let
\begin{equation}\label{Eq:Ham_degGar}
	\begin{split}
		H_1 &= -\frac{1}{t_1}\det A_{N+2}
		- \sum_{j=2}^{N+1}\,\frac{\mathrm{tr}A_1A_j-t_1\theta_j}{t_1t_j},\\
		H_i &= \frac{\mathrm{tr}A_iA_1-t_1\theta_i}{t_i^2}
		+ \sum_{j=2,j\neq i}^{N+2}
		\frac{\mathrm{tr}A_iA_j-\theta_i\theta_j}{t_i-t_j}\quad (i=2,\ldots,N).
	\end{split}
\end{equation}
\end{prop}
Proposition \ref{Prop:1-form_degGar} allows us to define the $\tau$-function
$\tau_0=\tau_0(t)$ by
\begin{equation}
	d\log\tau_0 = \omega_0,
\end{equation}
up to multiplicative constants.

Let $L_2$ be a subset of $\mathbb{Z}^{N+2}$ defined as
\begin{equation}
	L_2 = \left\{\nu=(\nu_2,\ldots,\nu_{N+3})\in\mathbb{Z}^{N+2}\bigm|
	|\nu|=\nu_2+\cdots+\nu_{N+3}\in2\,\mathbb{Z}\right\}.
\end{equation}
Then $S(1,\ldots,1,2;N)$ is invariant under the action of the Schlesinger
transformations $T_{\nu}$ $(\nu\in L_2)$ which act on the parameters as
follows (see \cite{JM}):
\begin{equation}
	T_{\nu}\,(\theta_j)=\theta_j+\nu_j\quad (j=2,\ldots,N+3).
\end{equation}
We give explicitly the action of $T_{\nu}$ on the dependent variables in
Appendix \ref{Sec:Sch_Trf}.

Let us define a family of $\tau$-functions by
\begin{equation}\label{Eq:Tau_degSch}
	d\log\tau_{\nu} = T_{\nu}\,(\omega_0)\quad (\nu\in L_2).
\end{equation}
\begin{rem}\rm
A family of $\tau$-functions for $S(1,\ldots,1,2;N)$ can be identified with
that for $G(1,\ldots,1,2;N)$ by
\begin{equation}
	\sum_{i=1}^{N}\,K_i\,ds_i = T_{(0,\ldots,0,1,0,-1)}\,(\omega_0).
\end{equation}
\end{rem}

Conversely we can express a solution of $S(1,\ldots,1,2;N)$ in terms of
$\tau$-functions as follows.
By
\begin{equation}
	T_{(0,\ldots,0,2)}\,(H_i) = H_i+D_i\log b_{\infty}\quad (i=1,\ldots,N),
\end{equation}
where $D_i=\partial/\partial t_i$,
we obtain
\begin{prop}
A solution of $S(1,\ldots,1,2;N)$ is expressed by means of $\tau$-functions as
follows{\rm:}
\begin{equation}
	\begin{split}
		&a_1 = \frac{t_1}{\theta_{N+3}}
		\left(D_1D_{N+3}\log\tau_0-\rho\right),\quad
		b_1 = t_1D_1\frac{\tau_{(0,\ldots,0,2)}}{\tau_0},\\
		&a_i = \frac{1}{\theta_{N+3}}\,
		\left(D_iD_{N+3}\log\tau_0-\theta_i\rho\right),\quad
		b_i = D_i\frac{\tau_{(0,\ldots,0,2)}}{\tau_0}\quad
		(i=2,\ldots,N),\\
		&a_{N+1} = \frac{1}{\theta_{N+3}}\,\biggl\{(D_{N+1}+1)D_{N+3}\log\tau_0
		- \rho(\rho+\theta_{N+1}+\theta_{N+3})\biggr\},\\
		&b_{N+1}
		= (D_{N+1}+\theta_{N+3}+1)\,\frac{\tau_{(0,\ldots,0,2)}}{\tau_0},\\
		&a_{N+2} = \frac{1}{\theta_{N+3}}\,\biggl\{(D_{N+2}-1)D_{N+3}\log\tau_0
		- \rho(\rho+\theta_{N+2}+\theta_{N+3})\biggr\},\\
		&b_{N+2}
		= (D_{N+2}-\theta_{N+3}-1)\,\frac{\tau_{(0,\ldots,0,2)}}{\tau_0},
	\end{split}
\end{equation}
where
\begin{equation}
	\begin{split}
		&D_{N+1} = -\sum_{i=1}^{N}\,t_iD_i,\quad
		D_{N+2} = t_1D_1+\sum_{j=2}^{N}\,(t_j-1)D_j,\\
		&D_{N+3}
		= -t_1D_1+\sum_{i=2}^{N}\,t_i(t_i-1)D_i.
	\end{split}
\end{equation}
\end{prop}

%% Section 2.3
\subsection{Coalescence structures}
As is known in \cite{IKSY}, the Garnier system $G(1,\ldots,1;N)$ is equivalent
to the Schlesinger system, denoted by $S(1,\ldots,1;N)$:
\begin{equation}\label{Sys:Sch}
	dA_j=\sum_{i=1,i\neq j}^{N+2}[A_i,A_j]\,d\log\,(t_j-t_i),\quad
	(j=1,\ldots,N+2),
\end{equation}
with the following conditions:
\begin{enumerate}
\item\quad $\mathrm{det}A_j=0$,\quad
$\mathrm{tr}A_j=\theta_j\notin\mathbb{Z}$\quad $(j=1,\ldots,N+2)$;
\item\quad The matrices $A_j$ satisfy
\begin{equation}
	A_{\infty}:=-\sum_{j=1}^{N+2}\,A_j
	= \begin{pmatrix}\rho&0\\0&\rho+\theta_{N+3}\end{pmatrix},\quad
	\theta_{N+3}\notin\mathbb{Z}.
\end{equation}
\end{enumerate}

Let $L_1$ be a subset of $\mathbb{Z}^{N+3}$ defined as
\begin{equation}
	L_1 = \left\{\mu=(\mu_1,\ldots,\mu_{N+3})\in\mathbb{Z}^{N+3}\bigm|
	|\mu|=\mu_1+\cdots+\mu_{N+3}\in2\,\mathbb{Z}\right\}.
\end{equation}
Then a family of $\tau$-functions for $S(1,\ldots,1;N)$ is defined by
\begin{equation}\label{Eq:Tau_Sch}
	d\log\tau_{\mu} = \sum_{i=1}^{N}\sum_{j=1,j\neq i}^{N+2}
	\frac{1}{t_i-t_j}\,T_{\mu}\,(\mathrm{tr}A_iA_j-\theta_i\theta_j)\,dt_i\quad
	(\mu\in L_1).
\end{equation}
Here we let $T_{\mu}$ be the Schlesinger transformations given in \cite{SUZ}.

The system $S(1,\ldots,1,2;N)$ is obtained from $S(1,\ldots,1;N)$ by the replacement
\begin{equation}\label{Eq:Sch2degSch}
	\begin{array}{lll}
		\theta_1\to1/\vep,& \theta_{N+2}\to\theta_{N+2}-1/\vep,&
		t_1\to\vep t_1,\\[4pt]
		\ds A_1\to\frac{A_1}{\vep t_1},&
		\ds A_{N+2}\to A_{N+2}-\frac{A_1}{\vep t_1},
	\end{array}
\end{equation}
and taking a limit $\vep\to0$.
Then \eqref{Eq:Tau_Sch} is also transformed into \eqref{Eq:Tau_degSch} via
\begin{equation}
	\tau_{\mu}\to\tau_{\nu}\quad (\mu\in L_1),
\end{equation}
where
\begin{equation}
	\nu=(\mu_2,\ldots,\mu_{N+1},\mu_1+\mu_{N+2},\mu_{N+3})\in L_2.
\end{equation}

%% Section 3
\section{Classical transcendental solutions}
In this section, a family of classical transcendental solutions is presented.
This is reduced to a family of rational solutions expressed in terms of the
Schur polynomials.

We recall the definition of the Lauricella hypergeometric series $F_D$.
For each $m=(m_1,\ldots,m_N)$, we let
\begin{equation}
	t^m = t_1^{m_1}\,\ldots\,t_N^{m_N},\quad |m| = m_1+\cdots+m_N.
\end{equation}
The series $F_D$ is defined by
\begin{equation}
	F_D(\alpha,\beta_1,\ldots,\beta_N,\gamma;t)
	= \sum_{m\in(\mathbb{Z}_{\geq0})^{N}}
	\frac{(\alpha)_{|m|}(\beta_1)_{m_1}\ldots(\beta_N)_{m_N}}
	{(\gamma)_{|m|}(1)_{m_1}\ldots(1)_{m_N}}\,t^m,
\end{equation}
where
\begin{equation}
	(\alpha)_k = \alpha(\alpha+1)\ldots(\alpha+k-1).
\end{equation}

Via \eqref{Eq:Sch2degSch} and taking a limit $\vep\to0$, $F_D$ is transformed
into
\begin{equation}\label{Eq:degHGF}
	\Phi_D(\alpha,\beta_2,\ldots,\beta_N,\gamma;t)
	=\sum_{m\in(\mathbb{Z}_{\geq0})^{N}}
	\frac{(\alpha)_{|m|}(\beta_2)_{m_2}\ldots(\beta_N)_{m_N}}
	{(\gamma)_{|m|}(1)_{m_1}\ldots(1)_{m_N}}\,t^m.
\end{equation}
We note that the series \eqref{Eq:degHGF} is a generalization of the
hypergeometric series $\Phi_1$ given by J. Horns (\cite{ERD}).

It is known that $S(1,\ldots,1;N)$ admits a family of solutions expressed by
$F_D$.
Let $\sigma^{(1)}_{m,n}$ $(m,n\in\mathbb{Z}_{\geq0})$ be functions defined as
follows:
\begin{equation}
	\begin{split}
		\sigma^{(1)}_{0,n} &= 1,\\
		\sigma^{(1)}_{1,n} &= (\theta_{N+2}-n)(\theta_{N+3}+n)\,
		t_1(1-t_1)^{-(\theta_{N+2}+\theta_{N+3}+1)}\\
		&\qquad\times F_D(-\theta_{N+3}-n,\theta_1,\ldots,\theta_N,
		-\theta_{N+1}-\theta_{N+3}-n+1;t).
	\end{split}
\end{equation}
and
\begin{equation}
	\sigma^{(1)}_{m,n} = \det\left(X^{i-1}Y^{j-1}\,
	\sigma^{(1)}_{1,n}\right)_{i,j=1,\ldots,m}\quad (m\geq2),
\end{equation}
where
\begin{equation}
	X = \frac{t_1}{t_1-1}\,\sum_{i=1}^{N}\,(t_i-1)D_i,\quad
	Y = \frac{1}{t_1-1}\,\sum_{i=1}^{N}\,t_i(t_i-1)D_i.
\end{equation}
\begin{thm}[\rm\cite{TSU2}\bf]
Let
\begin{equation}
	\tau_{(0,\ldots,0,m-n,m+n)} = C_{m,n}^{(1)}\,\sigma^{(1)}_{m,n}\quad
	(m,n\in\mathbb{Z}_{\geq0}),
\end{equation}
where
\begin{equation}
	C^{(1)}_{m,n} = t_1^{-m(m+1)/2}(1-t_1)^{m(\theta_{N+2}+\theta_{N+3}+m)}\,
	\prod_{k=1}^{m}\,\frac{1}{(\theta_{N+2}-n)_k}.
\end{equation}
When $\rho=0$, this is a family of $\tau$-functions for $S(1,\ldots,1;N)$.
\end{thm}

Via \eqref{Eq:Sch2degSch} and taking a limit $\vep\to0$, each
$\sigma^{(1)}_{m,n}$ is transformed into the function $\sigma^{(2)}_{m,n}$
defined as follows:
\begin{equation}
	\begin{split}
		\sigma^{(2)}_{0,n} &= 1,\\
		\sigma^{(2)}_{1,n} &= (\theta_{N+3}+n)\,t_1e^{-t_1}\\
		&\qquad\times \Phi_D(-\theta_{N+3}-n,\theta_2,\ldots,\theta_N,
		-\theta_{N+1}-\theta_{N+3}-n+1;t),
	\end{split}
\end{equation}
and
\begin{equation}
	\sigma^{(2)}_{m,n} = \det\left((t_1D_1)^{i-1}D_{N+3}^{j-1}\,
	\sigma^{(2)}_{1,n}\right)_{i,j=1,\ldots,m}\quad (m\geq2).
\end{equation}
Thus we obtain the following theorem.
\begin{thm}\label{Thm:CTSol}
Let
\begin{equation}\label{Eq:CTSol}
	\tau_{(0,\ldots,0,m-n,m+n)} = C^{(2)}_{m,n}\,\sigma^{(2)}_{m,n}\quad
	(m,n\in\mathbb{Z}_{\geq0}),
\end{equation}
where
\begin{equation}
	C^{(2)}_{m,n} = t_1^{-m(m+1)/2}e^{mt_1}.
\end{equation}
When $\rho=0$, this is a family of $\tau$-functions for $S(1,\ldots,1,2;N)$.
\end{thm}

Recall the definition of the Schur polynomials.
For each partition $\lambda = (\lambda_1,\ldots,\lambda_l)$, the Schur
polynomial is a polynomial in $x=(x_1,x_2,\ldots)$ defined by
\begin{equation}
	S_{\lambda}(x) = \det\,\Bigl(p_{\lambda_i-i+j}(x)\Bigr)_{i,j=1,\ldots,l},
\end{equation}
where $p_n(x)$ are the polynomials defined as
\begin{equation}\label{Eq:Spe_Poly}
	p_n(x) = \sum_{k_1+2k_2+\ldots+nk_n=n}
	\frac{x_1^{k_1}x_2^{k_2}\ldots x_n^{k_n}}{k_1!k_2!\ldots k_n!}.
\end{equation}

In a similar manner as \cite{TSU2}, the $\tau$-functions given by
\eqref{Eq:CTSol} are reduced to those expressed in terms of the Schur
polynomials.
\begin{thm}\label{Thm:RatSol}
Let
\begin{equation}
	\tau_{(0,\ldots,0,m-n,m+n)} = S_{(n^m)}(x)\quad
	(m,n\in\mathbb{Z}_{\geq0}),
\end{equation}
where we use the notation $(n^m)=(n,\ldots,n)$ and let
\begin{equation}
	x_1 = t_1 + \sum_{j=2}^{N+1}\,t_j\,\theta_j,\quad
	x_k = \frac{1}{k}\,\sum_{j=2}^{N+1}\,t_j^k\,\theta_j\quad (k\geq2).
\end{equation}
When $\rho=\theta_{N+3}=0$, this is a family of $\tau$-functions for
$S(1,\ldots,1,2;N)$.
\end{thm}

%% Section 4
\section{Algebraic solutions}
In this section, we present a family of algebraic solutions expressed in terms of the universal characters.

We recall the definition of the universal character introduced by K. Koike
\cite{KOI}, which is a generalization of the Schur polynomial.
For each pair of partitions
$[\lambda,\mu]=[(\lambda_1,\ldots\lambda_l),(\mu_1,\ldots\mu_{l'})]$, the
universal character $S_{[\lambda,\mu]}(x,y)$ is defined as follows:
\begin{equation}
	S_{[\lambda,\mu]}(x,y) = \det\left(\begin{array}{ll}
		p_{\lambda_{l'-i+j}+i-j}(y),& 1\leq i\leq l'\\
		p_{\lambda_{-l'+i}-i+j}(x),& l'+1\leq i\leq l+l'\\
	\end{array}\right)_{1\leq i,j\leq l+l'},
\end{equation}
where $p_n(x)$ is the polynomial defined by \eqref{Eq:Spe_Poly}.

The system $S(1,\ldots,1;N)$ admits a family of solutions expressed in terms of
the universal characters.
Let
\begin{equation}
	\xi_i^2=1-t_i\quad (i=1,\ldots,N).
\end{equation}
\begin{thm}[\rm\cite{TSU1,TSU3}\bf]\label{Thm:Alg_Sol_Gar}
Let
\begin{equation}
	\tau_{(0,\ldots,0,m-n,0,m+n)} = N^{(1)}_{m,n}\,S_{[u!,v!]}(x,y)\quad
	(m,n\in\mathbb{Z}),
\end{equation}
where
\begin{equation}
	x_k = \frac{1}{k}
	\left(\theta_{N+2}+\sum_{i=1}^{N}\,\theta_i\,\xi_i^k\right),\quad
	y_k = \frac{1}{k}
	\left(\theta_{N+2}+\sum_{i=1}^{N}\,\theta_i\,\xi_i^{-k}\right),
\end{equation}
and
\begin{equation}
	\begin{split}
		&[u!,v!]=[(u,u-1,\ldots,1),(v,v-1,\ldots,1)],\\
		&u=|m+n-1/2|-1/2,\quad v=|m-n+1/2|-1/2.
	\end{split}
\end{equation}
When $\theta_{N+1}=1/2$ and $\theta_{N+3}=-1/2$, this is a family of
$\tau$-functions for $S(1,\ldots,1;N)$.
\end{thm}
Here we let
\begin{equation}
	N^{(1)}_{m,n}
	= \prod_{i=1}^{N}\,\xi_i^{-\theta_i(\theta_i+2m-2n+1)/2}
	\prod_{i=1}^{N}\,\left(\frac{\xi_i+1}{2}\right)^{-\theta_i\theta_{N+2}}\,
	\prod_{i,j=1,i<j}^{N}\left(\frac{\xi_i+\xi_j}{2}\right)^{-\theta_i\theta_j}.
\end{equation}

Via \eqref{Eq:Sch2degSch} and taking a limit $\vep\to0$, we obtain from Theorem
\ref{Thm:Alg_Sol_Gar} the following theorem.
\begin{thm}\label{Thm:AlgSol}
Let
\begin{equation}
	\tau_{(0,\ldots,0,m-n,0,m+n)} = N^{(2)}_{m,n}\,S_{[u!,v!]}(x,y),
\end{equation}
where
\begin{equation}
	x_k = \frac{1}{k}\left(\theta_{N+2}-\frac{k}{2}\,t_1
	+\sum_{i=2}^{N}\,\theta_i\,\xi_i^k\right),\quad
	y_k = \frac{1}{k}\left(\theta_{N+2}+\frac{k}{2}\,t_1
	+\sum_{i=2}^{N}\,\theta_i\,\xi_i^{-k}\right).
\end{equation}
When $\theta_{N+1}=1/2$ and $\theta_{N+3}=-1/2$, this is a family of
$\tau$-functions for $S(1,\ldots,1,2;N)$.
\end{thm}
Here we let
\begin{equation}
	\begin{split}
		N^{(2)}_{m,n} &= e^{\Delta_{m,n}}
		\prod_{i=2}^{N}\,\xi_i^{-\theta_i(\theta_i+2m-2n+1)/2}\\
		&\qquad\times
		\prod_{i=2}^{N}\,\left(\frac{\xi_i+1}{2}\right)^{-\theta_i\theta_{N+2}}
		\prod_{i,j=2,i<j}^{N}
		\left(\frac{\xi_i+\xi_j}{2}\right)^{-\theta_i\theta_j},
	\end{split}
\end{equation}
where
\begin{equation}
	\Delta_{m,n} = \frac{t_1^2}{32}+\frac{t_1}{4}\left(2m-2n+1
	+\theta_{N+2}+\sum_{i=2}^{N}\,\frac{2\,\theta_i}{1+\xi_i}\right).
\end{equation}

%% Appendix
\appendix

%% Appendix A
\section{Schlesinger transfromations}\label{Sec:Sch_Trf}
In this Appendix, we describe the action of the Schlesinger transformations for
$S(1,\ldots,1,2;N)$ on the dependent variables.

The group of the Schlesinger transformations $T_{\nu}$ $(\nu\in L_2)$ is
generated by the transformations
\begin{equation}
	\begin{split}
		T_1 &= T_{(0,\ldots,0,1,1)},\\
		T_2 &= T_{(-1,0\ldots,0,1)},\\
		T_3 &= T_{(0,-1,0,\ldots,0,1)},\\
		&\qquad\vdots\\
		T_{N+2} &= T_{(0,\ldots,0,-1,1)}.
	\end{split}
\end{equation}
The action of $T_k$ $(k=1,\ldots,N+2)$ on the dependent variables is described
as follows:
\begin{equation}
	\begin{split}
		T_1\,(A_1)
		&= R_2^{(1)}A_1E_2 + E_1A_1R_1^{(1)} - R_2^{(1)}A_{N+2}R_1^{(1)},\\
		T_1\,(A_{N+2})
		&= R_2^{(1)}A_{N+2}E_2 + E_1A_{N+2}R_1^{(1)} - E_1A_1E_2
		+ E_1R_1^{(1)}\\
		&\qquad + \sum_{i=2}^{N+1}\,\frac{1}{t_i}\,R_2^{(1)}A_iR_1^{(1)},\\
		T_1\,(A_j)
		&= R_2^{(1)}A_jE_2 + E_1A_jR_1^{(1)} - t_jE_1A_jE_2\\
		&\qquad - \frac{1}{t_j}\,R_2^{(1)}A_jR_1^{(1)}\quad (j=2,\ldots,N+1),
	\end{split}
\end{equation}
where
\begin{equation}
	\begin{split}
		&R_1^{(1)} = \frac{1}{(\theta_{N+3}+1)\,b_1}
		\left(\begin{array}{@{}c@{}}b_1\\d_1\end{array}\right)
		\left(\begin{array}{@{}cc@{}}\theta_{N+3}+1&b_{\infty}\end{array}
		\right),\\
		&R_2^{(1)} = \frac{1}{(\theta_{N+3}+1)\,b_1}
		\left(\begin{array}{@{}c@{}}-b_{\infty}\\\theta_{N+3}+1\end{array}
		\right)\left(\begin{array}{@{}cc@{}}-d_1&b_1\end{array}\right),
	\end{split}
\end{equation}
for $k=1$;
\begin{equation}
	\begin{split}
		T_k\,(A_1) &= E_1A_1R_1^{(k)} + R_2^{(k)}A_1E_2 + t_kE_1A_1E_2
		+ \frac{1}{t_k}\,R_2^{(k)}A_1R_1^{(k)},\\
		T_k\,(A_{N+2}) &= E_1A_{N+2}R_1^{(k)} + R_2^{(k)}A_{N+2}E_2
		+ t_kE_1A_{N+2}E_2 - E_1A_1E_2\\
		&\qquad + \frac{1}{t_k}\,R_2^{(k)}A_{N+2}R_1^{(k)}
		+ \frac{1}{t_k^2}\,R_2^{(k)}A_1R_1^{(k)},\\
		T_k\,(A_k) &= E_1A_kR_1^{(k)} + R_2^{(k)}A_kE_2 - R_2^{(k)}E_2
		- \frac{1}{t_k^2}\,R_2^{(k)}A_1R_1^{(k)}\\
		&\qquad - \sum_{i=2,i\neq k}^{N+2}
		\frac{1}{t_k-t_i}\,R_2^{(k)}A_iR_1^{(k)},\\
		T_k\,(A_j) &= E_1A_jR_1^{(k)} + R_2^{(k)}A_jE_2 + (t_k-t_j)\,E_1A_jE_2\\
		&\qquad + \frac{1}{t_k-t_j}\,R_2^{(k)}A_jR_1^{(k)}\quad
		(j\neq 1,k,N+2),
	\end{split}
\end{equation}
where
\begin{equation}
	\begin{split}
		&R_1^{(k)} = \frac{1}{(\theta_{N+3}+1)\,b_k}
		\left(\begin{array}{@{}c@{}}b_k\\-a_k\end{array}\right)
		\left(\begin{array}{@{}cc@{}}\theta_{N+3}+1&b_{\infty}\end{array}
		\right),\\
		&R_2^{(k)} = \frac{1}{(\theta_{N+3}+1)\,b_k}
		\left(\begin{array}{@{}c@{}}-b_{\infty}\\\theta_{N+3}+1\end{array}
		\right)\left(\begin{array}{@{}cc@{}}b_k&a_k\end{array}\right),
	\end{split}
\end{equation}
for $k=2,\ldots,N+1$;
\begin{equation}
	\begin{split}
		T_1\,(A_1) &= E_1A_1R_1^{(N+2)} + R_2^{(N+2)}A_1E_2
		- R_2^{(N+2)}A_{N+2}R_1^{(N+2)},\\
		T_1\,(A_{N+2}) &= E_1A_{N+2}R_1^{(N+2)} + R_2^{(N+2)}A_{N+2}E_2 
		- E_1A_1E_2 - R_2^{(N+2)}E_2\\
		&\qquad + \sum_{i=2}^{N+1}\,\frac{1}{t_i}\,R_2^{(N+2)}A_iR_1^{(N+2)},\\
		T_1\,(A_j)
		&= E_1A_jR_1^{(N+2)} + R_2^{(N+2)}A_jE_2 - t_jE_1A_jE_2\\
		&\qquad - \frac{1}{t_j}\,R_2^{(N+2)}A_jR_1^{(N+2)}\quad
		(j=2,\ldots,N+1),
	\end{split}
\end{equation}
where
\begin{equation}
	\begin{split}
		&R_1^{(N+2)} = \frac{1}{(\theta_{N+3}+1)\,b_1}
		\left(\begin{array}{@{}c@{}}b_1\\-a_1\end{array}\right)
		\left(\begin{array}{@{}cc@{}}\theta_{N+3}+1&b_{\infty}\end{array}
		\right),\\
		&R_2^{(N+2)} = \frac{1}{(\theta_{N+3}+1)\,b_1}
		\left(\begin{array}{@{}c@{}}-b_{\infty}\\\theta_{N+3}+1\end{array}
		\right)\left(\begin{array}{@{}cc@{}}b_1&a_1\end{array}\right),
	\end{split}
\end{equation}
for $k=N+2$.\\

%% Acknowledgement
{\bf Acknowledgement}\quad
The auther is grateful to Professor Masatoshi Noumi and Dr. Teruhisa Tsuda for
valuable discussions and advices.

%% Reference


\begin{thebibliography}{9}
\bibitem{ERD}
	A. Erderyi et al.,
	Higher transcendental functions,
	MacGraw-Hill (1953).
\bibitem{IKSY}
	K. Iwasaki, H. Kimura, S. Shimomura and M. Yoshida,
	From Gauss to Painlev\'{e} --- A Modern Theory of Special Functions,
	Aspects of Mathematics \textbf{E16} (Vieweg, 1991).
\bibitem{JMU}
	M. Jimbo, T.Miwa and K.Ueno,
	Monodromy preserving deformation of linear ordinary differential equations
	with rational coefficients I,
	Physica \textbf{2D} (1981), 306-352.
\bibitem{JM}
	M. Jimbo and T.Miwa,
	Monodromy preserving deformation of linear ordinary differential equations
	with rational coefficients II,
	Physica \textbf{2D} (1981), 407-448.
\bibitem{KAW1}
	H. Kawamuko,
	On the holonomic deformation of linear differential equations,
	Proc. Japan Acad. Ser. A Math. Sci. \textbf{73} (1997), 152-154.
\bibitem{KAW2}
	H. Kawamuko,
	On the polynomial Hamiltonian structure associated with the $L(1,g+2;g)$
	type,
	Proc. Japan Acad. Ser. A Math. Sci. \textbf{73} (1997), 155-157.
\bibitem{KIM}
	H. Kimura,
	The degeneration of the two dimentional Garnier system and the Polynomial
	Hamiltonian structure,
	Ann. Mat. Pura Appl., \textbf{155} (1989), 25-57.
\bibitem{KO}
	H. Kimura and K. Okamoto,
	On particular solutions of the Garnier systems and the hyper geometric
	functions of several variables,
	Quarterly J. Math., \textbf{37} (1986), 61-80.
\bibitem{KOI}
	K. Koike,
	On the decomposition of tensor products of the representations of the
	classical groups: By means of the universal characters,
	Adv. Math., \textbf{74} (1989), 57-86.
\bibitem{LIU}
	D. Liu,
	Holonomic deformation of linear differential equations of $A_g$ type and
	polynomial Hamiltonian structure,
	{\it Ph.D thesis}, (Univ. Tokyo, 1997).
\bibitem{MSD1}
	T. Masuda,
	On a class of algebraic solutions to the Painlev\'{e} VI equation, its
	determinant formula and coalescence cascade,
	Funkcial. Ekvac. \textbf{46} (2003), 121-171.
\bibitem{MSD2}
	T. Masuda,
	Classical transcendental solutions of the Painlev\'{e} equations and their
	degeneration,
	to appear in Tohoku Math. J. \textbf{56} (2004), nlin-SI/0302026.
\bibitem{OKM1}
	K. Okamoto,
	Isomonodromic deformation and Painlev\'{e} equations, and the Garnier
	system,
	J. Fac. Sci. Univ. Tokyo Sect. IA, Math. \textbf{33} (1986), 575-618.
\bibitem{OKM2}
	K. Okamoto,
	Studies on the Painlev\'{e} equations II,
	Japan. J. Math. \textbf{13} (1987), 47-76.
\bibitem{OKM3}
	K. Okamoto,
	The Hamiltonians associated with the Painlev\'{e} equations,
	The Painlev\'{e} property: One Century Later, ed. R. Conte, CRM Series in
	Mathematical Physics, (Springer, 1999).
\bibitem{SUZ}
	T. Suzuki,
	Affine Weyl group symmetry of the Garnier system,
	{\it preprint}, math-ph/0312068.
\bibitem{TSU1}
	T. Tsuda,
	Universal characters and Integrable systems,
	{\it Ph.D thesis}, (Univ. Tokyo, 2003).
\bibitem{TSU2}
	T. Tsuda,
	Rational solutions of the Garnier system in terms of Schur polynomials,
	Int. Math. Res. Not. \textbf{43} (2003), 2341-2358.
\bibitem{TSU3}
	T. Tsuda,
	Toda equations and special polynomials associated with the Garnier system,
	submitted.
\end{thebibliography}
\end{document}